\documentclass{article}

\usepackage[text={6in,9in},foot=0.6in]{geometry}
\usepackage{verbatim}

\usepackage{amsmath, amssymb, amsthm}
\usepackage{graphicx}
\usepackage{amsfonts}
\usepackage{graphicx} \usepackage{amsmath}

\newcommand{\tp}{\tilde{p}}

\newcommand{\beq}{\begin{equation}}
\newcommand{\eeq}{\end{equation}}

\newcommand{\vp}{\varphi}
\newcommand{\al}{\alpha}
\newcommand{\de}{\delta}
\newcommand{\ga}{\gamma}
\newcommand{\be}{\beta}
\newcommand{\ep}{\varepsilon}
\newcommand{\la}{\lambda}

\newcommand{\bH}{\mathbf{H}}
\newcommand{\bQ}{\mathbf{Q}}

\newcommand{\bw}{\mathbf{w}}

\newcommand{\avt}{\langle t\rangle}

\newcommand{\rg}{\rangle}
\newcommand{\lgl}{\langle}

\newcommand{\e}{{\rm e}}
\newcommand{\deq}{\stackrel{\cdot}{=}}

\begin{document}

\title{Single-Seed Cascades on Clustered Networks}
\author{John K. McSweeney\\
Rose-Hulman Institute of Technology}
\maketitle

\begin{abstract}
We consider a dynamic network cascade process developed by Watts \cite{W} applied to a random networks with a specified amount of clustering, belonging to a class of random networks developed by Newman \cite{New09}. We adapt tree-based methods from \cite{W} to formulate an appropriate two-type branching process to describe the spread of a cascade started with a {\it single} active node, and obtain a fixed-point equation to implicitly express the extinction probability of such a cascade. In so doing, we also recover a special case of a formula of Hackett et al. \cite{HMG} giving conditions for certain extinction of the cascade. 

\end{abstract}

\section{Introduction}


\subsection{Motivation}

Watts \cite{W} describes a process that can be thought to model the spread of a trend on a social network -- individuals will adopt a trend if a sufficiently high {\it proportion} of their friends do. The interpretation of such dynamics has been carried over to other network settings -- spread of failures through a power grid \cite{DCM,KTM}, cascading failures of financial institutions \cite{Hurd}, adoption of new technology \cite{Ar}, or even solutions of crossword puzzles \cite{mcsweeney}. From these local contact rules we wish to answer global questions about the spread of a phenomenon on a network, especially:  for a given class of random networks, with what probability does a trend started with a small number of adherents `explode' and propagate through a nontrivial proportion of the whole network?

\subsection{Description of the Model}

In honor of \cite{W}, we shall refer to the following as {\it Watts' Cascade Process}. Consider a simple, undirected network (graph) $G=(V,E)$ where $V$ is the set of nodes (vertices) and $E$ the set of edges (links). Assign a threshold $\vp_x \in [0,1]$ to each node $x\in V$, where the $\vp_x$ are independent and identically distributed (i.i.d.) random variables which remain fixed throughout the process. We will often consider the case where all thresholds are equal, $\vp_x \equiv \vp$ for all $x$, which can be considered a random distribution consisting of a point mass at $\vp$. At any time, each node is either in state {\it active} (A) or {\it inactive} (I). The process is initially `seeded' with some nonempty set of active nodes, and then any inactive node $x$ becomes active if and only if  the {\it proportion} of its neighbors in the network that are active reaches a level greater than or equal to $\vp_x$. Once a node becomes active, it is never allowed to deactivate; the order and/or rate at which these activations are performed are irrelevant for our purposes. 

The choice of a model which considers the {\it proportion} of active neighbors as the criterion for activation is a sensible modeling choice for scenarios where each agent (node) only has a finite capacity for absorbing information, and the more relationships (edges) it has, the less significant each one is. For example, in terms of adoption of a trend or new technology, a user for whom 4 out of 5 friends have adopted a particular technology is more likely to be swayed than one for whom 6 out of 50 friends have. 

Watts' cascade process is a special case of a class of processes that are sometimes referred to as (network) ``diffusions'' \cite{A}; it is similar to bootstrap percolation \cite{Adler}, but differs in that here it is the {\it proportion} and not the {\it absolute number} of active neighbors that determines a node's activation. Hackett et al. \cite{GMH} discuss a large class of network-based dynamical processes that includes Watts' model but also other `monotone' processes (where the state of any node may change at most once throughout the process) such as site percolation, bond percolation, and $k$-core sizes. The unifying quantity is what they call a {\it response function} $F(m,k)$: the probability that a node of degree $k$ will be activated by $m$ active neighbors. For Watts' model we have $F(m,k)=C(m/k)$, where $C$ is the cumulative distribution function of the thresholds $\vp$ (a step function in the case of constant $\vp$, which we consider here). In this work, we will only consider Watts' process, which presents some unique challenges due to the interaction of the response function and the connectivity of the network -- reflected in the fact that $F(m,k)$ depends on $k$, which is not the case for bootstrap percolation, for example. The relation between the connectivity of the network and the propensity for cascade spread is very much nontrivial: on the one hand, more edges means more links along which the cascade can spread; however, the presence of many edges connected to a given node makes that node more resistant to activation. Indeed, the most highly-connected nodes are, all else being equal, the most resistant to activation. 

For each instance of this process, we define the {\it cascade} to be the final set of active nodes once there no more nodes can be activated. We say that a {\it global cascade} occurs if the cascade occupies an asymptotically nonzero fraction of the whole network, in the limit as the network size $n $ tends to $\infty$.  A natural question in such a scenario is: for a given class of random networks, what is the probability of triggering a global cascade starting from a `small' set of randomly chosen initially active seeds? 

\subsection{Previous Work}\label{sec:pw}

Watts' original work \cite{W}, and many subsequent articles \cite{GalCo,GC} considered cascades on classes of networks that were {\it locally tree-like}, in the sense that, as $n \to \infty$, for all nodes $x$, with probability $\to 1$, there is a function $\omega(n) \to \infty$ such that the induced subgraph consisting of all nodes at distance less than $\omega(n)$ from $x$ is a tree. The Molloy-Reed {\it configuration model} networks \cite{MR}, which are built from a specified degree sequence, form such a class.  The process considered in \cite{W} on locally tree-like networks involved a single active seed node, which was viewed as the root of its tree-like neighborhood to determine whether the cascade ``exploded'' as it spread down the tree. On the other hand, Gleeson et al. \cite{GC} considered networks that were still locally tree-like, but posited a small but nonzero fraction $\rho_0$ of initially active nodes, and then considered the tree-like neighborhood of an arbitrarily chosen node $r$. In this case, nodes at ``level $\infty$'' of the tree are independently active with probability $\rho_0$, and the probability of $r$ getting activated is the expected cascade size. Hackett et al. \cite{HMG} also considered this process on a class of networks described in \cite{Miller09} and \cite{New09} which extend the configuration model to allow a prescribed distribution of triangles as well, and thus are not locally tree-like. Similar techniques are applicable to these ``clustered'' networks, but the tree-like configuration must be amended to allow edges between siblings. One can then ask: how does the presence of clustering (short loops) in the network affect the probability of a single seed triggering a global cascade? 

\subsection{Focus of the Current Work}

\subsubsection{Single-Seed Cascades}
We focus on the following question: if the cascade process is started with a  {\it single} uniformly randomly chosen initially active seed $r$, and we consider a network with a high density of triangles (described below), what is the {\it probability} of triggering a global cascade? This differs from the analysis in \cite{HMG} in that we (as did Watts) consider the probability of random extinction of the cascade even in an environment that may be nevertheless ``supercritical'' for the spread of a cascade (with a natural interpretation in the context of branching process theory, which we shall quantify shortly). The analysis in \cite{HMG} involves small {\it relative} numbers of seed nodes, but a large {\it absolute} number of them: for example, their simulations use $100$ initially active seeds in a network of size $10^5$, so that the initially active seed fraction $\rho_0 = 10^2/10^5 = 10^{-3}$ is indeed quite small, while the absolute number of seeds is nevertheless large. If the network is in any way `supercritical' for the spread of a global cascade, then the probability that all 100 seeds will fail to do so is essentially zero. 

\subsubsection{Clustered Networks}\label{sec:networks}

The networks we consider here are a special case of so-called ``stubs-and-corners'' networks first described by Miller \cite{Miller09} and Newman \cite{New09}. For a network with $n$ nodes, there is a given bivariate probability distribution $\{p_{st}\}_{s,t\geq 0}$, where each node is independently assigned $s$ `stubs' and $t$ `corners' with probability $p_{st}$. The network is then formed by, uniformly at random, (a) pairing up the stubs to create single edges, and (b) grouping the corners by threes to create triangles in the network. (One first has to condition on the total number of stubs being even and the total number of corners being a multiple of three.)  The case $p_{st}=\delta_{t,0}p_s$ reduces to the well-known (locally tree-like) configuration-model network of Molloy and Reed \cite{MR}; we will refer to such networks as {\it unclustered}. In this paper we consider networks with triangles; in the interest of simplicity, we shall only consider $p_{st}=\delta_{s,0}\cdot p_t$ for some probability distribution $\{p_t\}_{t \geq 0}$ -- that is, no isolated edges are allowed; our analysis should extend to the case where $p_{st}>0$ for some $s>0$, at the cost of making the formulas much messier.  Thus $p_t$ is the probability that a randomly chosen node is part of $t$ triangles, and in particular that it has degree $2t$; we shall call $\{p_t\}$ the {\it triangle-distribution} of the network, and refer to such networks as {\it clustered}. We may have occasion to use the usual degree distribution, which we denote $\tp_k = p_{k/2}$. 

 
Such networks would be locally tree-like (in the sense defined in Section \ref{sec:pw}) if we were to disregard the triangles that are inherent to the model, meaning that for any given node $r$, its neighborhood  can be viewed as a tree structure but where each node has an edge linking it to one of its siblings (as well as to its parent).  We follow the terminology of Ikeda et al. \cite{IHN} and call this the {\it Random Cactus Layout} of the network rooted at $r$.  For the cactus layout rooted at $r$, we refer to each linked sibling pair as a {\it Node-Pair} (NP), and a NP that consists of two linked  children of some node $z$ will be called  a {\it Child-Pair} (CP) of $z$. The distribution of CP counts of the root $r$ is just $\{p_t\}$, but for a non-root node $v$, its CP count is distributed as $q_{t+1} = (t+1)p_{t+1}/\avt$, where $\{q_t\} := \{tp_t/\avt\}$ is the size-biased triangle distribution, in which $\langle t \rangle$ denotes the mean number of triangles incident to each node (see Durrett \cite{D}, for example). The idea behind this size-biased distribution is that the fact that a node is reached via an edge biases its distribution -- higher-degree nodes are likelier to be reached.   

The seed node for the process is used as the root of the random cactus, and we allow the cascade to percolate down the cactus. 
The probability that the cascade becomes global is the probability that the total number of NPs that contain one or two active nodes diverges; in the next section we formulate the appropriate branching process so as to obtain a condition for this event. For a node $v$ in a network, define $d(v)$ to be the degree (number of edges incident to) of $v$. For a locally tree-like network with a single initially active seed $r$, it is clear that, initially, one need only focus on the nodes that Watts calls {\it vulnerable}: nodes that can be activated by a single active neighbor, i.e. nodes $v$ where $\vp_v\leq 1/d(v)$. This is because a node can only get activated by a single node -- its parent -- in the local tree rooted at $r$ in the early phase of the process. However, for in the random cactus of a network with triangles, a node $v$ where $1/d(v) < \vp_v \leq 2/d(v)$ might also be deemed ``vulnerable'' in some sense: $v$'s parent may get activated, which may then activate $v$'s sibling, and the combined effect of these two will then be enough to activate $v$. The effect of such nodes will be illustrated in formulas such as \eqref{CascadeCondition} below.

\subsubsection{Limitations of the Model}

This approach does not address what the {\it distribution} of the final cascade size is, only the probability that it occupies at least a positive fraction of the network. We expect that any cascade that becomes global will actually spread to the entire network (or at least the connected component containing the root $r$): nodes that were not initially part of the cascade in the vicinity of $r$ may now have several neighbors which have become activated through loops in the network that appear away from the cactus-like neighborhood of $r$, and can now become activated themselves in a ``second wave'' of activation. 
As noted by Watts \cite{W}, it is an open question to prove a result about the distribution of the final cascade size. To obtain a final cascade size that is a fraction $\al \in (0,1)$ of the largest connected component, one would likely need one of the following situations: a highly ``modular'' network, where there are ``bottlenecks'' through which the cascade is unable to pass (see \cite{GalCo}, for instance), or e.g. a strongly varying distribution of cascade thresholds $\vp_x$, where the nodes with the higher threshold values never activate, but the network is sufficiently connected that these ``resistant'' nodes do not prevent the cascade from spreading through the low-threshold nodes. Both of these issues are present in \cite{mcsweeney} where the cascade process with variable thresholds is used to model the progression of the solution of a crossword puzzle.   

Our analysis is also sensitive to finite-size effects. As pointed out by Whitney \cite{Wh}, for the case of locally tree-like networks, this tree- (or cactus-) based approach is only valid in the theoretical $n\to \infty$ limit, and deviations from the asymptotic theory may appear even for large values of $n$. Indeed, we expect the diameter of many random network models to be approximately logarithmic in the size of the network, and thus for finite networks, loops will appear in a relatively small neighborhood of the root, which may invalidate the analysis based on a tree or cactus approximation. For example, in a tree-like network, even for seemingly very large networks (of size on the order of $10^4$), the presence of these loops may activate some nodes which require two active neighbors to activate, leading to a wider parameter range where cascades may happen; this accounts for the slight discrepancies between the numerics and the theory in Fig. 1 of \cite{W}, for example. Gleeson and Cahalane \cite{GC} have also noted that the probability of a global cascade is sensitively dependent on the initial seed fraction $\rho_0$ near 0. The reason for this is similar -- some nodes $v$ requiring two active neighbors to activate may acquire two such neighbors (presumably from different seeds), and will then become active themselves; this accounts for the presence of $C_2$ in Equation (6) of \cite{GC}.


\section{Model Analysis}

\subsection{Branching Process Formulation}

Using the seed node $r$ as the root of the random cactus, we set up a two-phase, two-type branching process for the cascade process seeded at $r$ as follows. The units in the branching process will be {\it Node-Pairs} (NPs), i.e. linked siblings in the cactus layout rooted at $r$, and each node has a set of  {\it Child-Pairs} (CPs). We say that an NP is of type $i$ iff it consists of $i$ active nodes, for $i\in\{0,1,2\}$; a global cascade occurs iff the sum of the number of Type 1 and 2 NPs diverges, or equivalently, the total size of the branching process restricted to nodes of Type 1 and 2 is not finite.  We call the branching process {\it two-phase} since the CP distribution for the root node is not the same as for internal nodes in the cactus; this is a common theme in percolation-type analyses on configuration-model networks (see \cite{D}, Chap. 3, for instance).  Here is how we set up the offspring distributions for the branching process.

Let $p_1(k,l)$ denote the distribution of CPs of Types $1$ and $2$ of the root node,
\[
p_1(k,l) = P(\mbox{root node has $k$ Type-1 and $l$ Type-2 CPs})
\]
and $q_i(k,l)$ the distribution of CP types of a NP of type $i$ from generation 1 or greater, for $i=1,2$: 
\[
q_i(k,l) = P(\mbox{a NP of type $i$ has $k$ Type-1 CPs and $l$ Type-2 CPs})~,~~~~~i=1,2.
\]
(We may safely ignore NPs of Type 0, since none of their descendants in the cactus can ever be activated in the initial activation phase down the cactus.) 
Let $P_1(x,y), Q_1(x,y)$ and $Q_2(x,y)$ be the bivariate generating functions for $p_1,q_1,$ and $q_2$ respectively:
\[ 
P_1(x,y) = \sum_{k,l} p_1(k,l)x^ky^l,~~~~~~~Q_i(x,y) = \sum_{k,l} q_i(k,l) x^k y^l,~~i=1,2
\]
and let $\bQ(x,y)=(Q_1(x,y),Q_2(x,y))$. Let $h_i(x,y),~i=1,2$ denote the generating function for the final counts of Type 1 and 2 NPs for a cactus where {\bf every} generation evolves according to $(q_1,q_2)$ (the homogeneous part), starting with a NP of type $i$:
\[
h_i(x,y) = \sum_{k,l} P(\mbox{Cactus seeded with a Type-$i$ NP has $k$ Type-1 and $l$ Type-2 NPs})~\cdot x^ky^l
\]
and let $\bH(x,y):=(h_1(x,y),h_2(x,y))$. We have the relations
\begin{align}\label{fph1}
h_1(x,y) &= x \sum_{k,l} q_1(k,l)h_1(x,y)^kh_2(x,y)^l\\ \label{fph2}
h_2(x,y) &= y \sum_{k,l} q_2(k,l)h_1(x,y)^kh_2(x,y)^l
\end{align}
where the factors of $x$ and $y$ in front of the summations are due to the fact that we're counting the root. This system \eqref{fph1}--\eqref{fph2} can be more succinctly written as 
\begin{equation}\label{fpq}
\bH(x,y) = (xQ_1(\bH(x,y)),yQ_2(\bH(x,y))) = (x,y) \odot \bQ(\bH(x,y)),
\end{equation}
where $\odot$ denotes elementwise vector multiplication.

Let $g_1(x,y)$ denote the generating function for the counts of NPs of types 1 and 2 in the total cactus, where in the first generation the CP counts are $p_1$-distributed, and in subsequent generations $(q_1,q_2)$-distributed -- the two-phase BP. We therefore have
\begin{equation}\label{g1}
g_1(x,y) = x\sum_{k,l} p_1(k,l) h_1(x,y)^kh_2(x,y)^l = xP_1(\bH(x,y)).
\end{equation}
The factor of $x$ in front of the sum is due to the fact that we pretend that the root node (the only node in the cactus without a sibling) is really a NP of type 1. Ultimately we are interested in $v:=g_1(1,1)$: the probability that the set of type-1 and -2 NPs has finite size, which is the probability that a cascade started at the root $r$ will {\bf not} become global. From \eqref{g1}, we have
\begin{equation}\label{totcasc}
v:=g_1(1,1) = P_1(\bw),
\end{equation}
where $\bw=\bH(1,1)$ is found (in theory) by finding the fixed point of \eqref{fpq}, i.e.  finding $\bw$ such that 
\begin{equation}\label{fp}
\bw = \bQ(\bw).
\end{equation}
The equation \eqref{fp} always has the trivial solution $\bw=(1,1)$, since $\bQ$ is a genuine bivariate probability generating function; further, basic branching process theory (see e.g. Feller \cite{F}) tells us that \eqref{fp} has a nontrivial solution iff 
\begin{equation}\label{cascond}
\mu_{11} \geq 1,~~~ \mu_{22}\geq 1 ~~~~\mbox{or}~~~~~ (1-\mu_{11})(1-\mu_{22}) \leq \mu_{12}\mu_{21},    
\end{equation}
where
\begin{equation}\label{mus}
\mu_{i1}:= \frac{\partial Q_i(1,1)}{\partial x},~~~~~~~  \mu_{i2}:= \frac{\partial Q_i(1,1)}{\partial y},~~~\mbox{ for }~~i=1,2.
\end{equation}
We note that $\mu_{ij}$ is the expected number of CPs of type $j$ from a parent of type $i$. This generalizes the well-known condition $\mu \leq 1$ for certain extinction of a 1-type branching process, where $\mu$ is the expected offspring size of an individual. (We shall see in fact that \eqref{cascond} becomes much simplified in the case we consider.) If a nontrivial $\bw$ is found (or estimated), it can be plugged into \eqref{totcasc} to find $v$, yielding a nonzero survival probability for the branching process.  


\subsection{Child-Pair Distributions by Parent Type}\label{sec:CP}

Finding formulas for $p_1(k,l)$ and $q_i(k,l)$ (and thus for their associated generating functions $P_1(x,y)$ and $\bQ(x,y)$) is not straightforward, since the state of a node 
affects its CP distribution; for example, the fact that a node has been activated indicates that it is likelier to have low degree. We note here that this issue does not arise for processes such as bootstrap percolation, where the response function $F(m,k)$ does not depend on $k$.

We first determine the distribution of the total number of CPs an NP of a given type has, regardless of the type of the CPs. Start with $Z_r$, the number of CPs of the root with an active parent. Since the root is automatically active,  the fact that it is active does not affect the distribution of $Z_r$ and so
\begin{equation}\label{p1t}
P(Z_r = t) = p_t,
\end{equation}
where $\{p_t\}_{t\geq 0}$ is the triangle-distribution defined in Section \ref{sec:networks}. For further generations, let $C_a$ denote the number of CPs of a node $a$ at a level $n \geq 1$ (without any conditioning on types); as mentioned in Section \ref{sec:networks}, $C_a$ has the size-biased distribution $\{q_t\}$,
\[
P(C_a = t) = q_{t+1} = (t+1)p_{t+1}/\avt.
\]
Unless stated otherwise, from this point forward we shall only consider uniform thresholds $\vp_x \equiv \vp$ for all $x$; the randomness in the process is therefore entirely due to the network topology and the choice of the seed node. Define $F_m(t):=F(m,2t)$ to be the probability that a node that is a member of $t$ triangles (i.e. has degree $2t$) can be activated by $m$ active neighbors. 
We define a node $x$ to be {\it $i$-vulnerable} if and only if $i$ active neighbors are sufficient to activate it, i.e. if $\vp_x\leq i/d(x)$.

Let $\al$ be the probability that a node $a$ in the cactus is 1-vulnerable; $\be$, the probability that it is 2- but not 1-vulnerable; and $\ga :=1-\al-\be$ the probability that it is not 2-vulnerable. 
Bearing in mind that a node $a$ with $C_a$ CPs in the cactus has total degree $2(C_a+1)$, we have the following expressions for $\al,\be,$ and $\ga$:
\begin{align} \label{albedefs1}
\al &= P\left(C_a+1\leq\frac{1}{2\vp}\right)~~~~~~~~ = \frac{1}{\avt} \sum_{t=1}^{1/2\vp} tp_t =\lgl tF_1 \rg/\avt,\\ \label{albedefs2}
\be &= P\left(\frac1{2\vp}<C_a+1\leq\frac1{\vp}\right) = \frac{1}{\avt} \sum_{t=1/2\vp}^{1/\vp} tp_t = \lgl t(F_2-F_1)\rg/\avt  \\ \label{albedefs3}
\ga &= P\left(\frac1{\vp}<C_a+1\right)~~~~~~~ = \frac{1}{\avt} \sum_{t=1/\vp}^\infty tp_t = \lgl (1-F_2)\rg/\avt.
\end{align}

{\bf Note:} To avoid ambiguities in the range of the above summations, we may assume throughout that $\vp$ is irrational. 

Let $Z_1$ and $Z_2$ denote the numbers of CPs of a Type 1 NP and a Type 2 NP respectively {\bf which have an active parent}, for generations $n\geq 1$, and let $r_1(k)$ and $r_2(k)$ denote their distributions, i.e. $P(Z_i=k) = r_i(k)$. 
Consider a Type-1 NP $(a,b)$ where $a$ is active and $b$ is inactive; this implies that $a$ was activated from its parent only and is thus 1-vulnerable (i.e. $\vp<1/d(a)$), but $b$ is not even $2$-vulnerable (since the activation of its parent and $a$ still does not activate it). In this case,   $\{r_1(t)\}_{t\geq 0}$ is then just the distribution of the number of CPs of $a$ under the above condition,
\begin{align}
\notag r_1(t)=P(Z_1=t) &= P\big(C_a=t\big|\mbox{$a$ is active and $b$ is inactive}\big) \\
\notag &= P\bigg(C_a=t\bigg|\frac{1}{2(C_a+1)}>\vp\bigg)\\
\notag & =\frac{P(C_a = t)}{\al},~~~~~t=0,1,2,\dots,\lfloor 1/2\vp -1\rfloor\\
& =\frac{(t+1)p_{t+1}}{\lgl tF_1 \rg},~~~~~t=0,1,2,\dots,\lfloor 1/2\vp -1\rfloor, \label{r1t}
\end{align}
and the mean of $Z_1$ is 
\begin{equation}\label{ez1}
E[Z_1] = \frac{\langle (t-1) t F_1\rangle }{\langle tF_1\rangle}.
\end{equation}

For $Z_2$, we have sum the CP counts of each active node in the pair, but these counts are dependent.  For both nodes $(a,b)$ in an NP to be active, one of the following disjoint events must hold:
\begin{enumerate}
\item $a$ was 1-vulnerable, and $b$ was 2-vulnerable (in which case $b$ may have needed both its parent and $a$ to activate);
\item $b$ was 1-vulnerable, and $a$ was 2- but not 1-vulnerable.
\end{enumerate}
Call these events 
\beq\label{ds}
D_1 := \{C_a+1<1/2\vp\}\cap \{C_b+1<2/2\vp\},~~~~~ D_2:=\{1/2\vp<C_a+1<2/2\vp\}\cap\{C_b+1<1/2\vp\},
\eeq
and let $D:=D_1 \cup D_2$. Note that $P(D_1) = \al^2 +\al\be$ and $P(D_2)= \al\be$. We have
\begin{align}\notag
r_2(t) &= P(C_a+C_b=t|D_1\cup D_2) = \frac{1}{P(D)}\left(P(C_a+C_b=t,D_1) + P(C_a+C_b=t,D_2)\right)\\
       &= \frac{1}{\al^2+2\al\be} \left(\sum_{{t_1,t_2: t_1+t_2=t,}\atop {{t_1<1/2\vp-1}\atop{t_2<1/\vp-1}}}q_{t_1+1}q_{t_2+1} + \sum_{{t_1,t_2: t_1+t_2=t,}\atop {{1/2\vp<t_1<1/\vp-1}\atop{t_2<1/2\vp-1}}}q_{t_1+1}q_{t_2+1}\right) \label{r2t}
\end{align}
and
\begin{align*}
E[Z_2] &= E[C_a+C_b| D_1 \cup D_2]\\
& =\frac{1}{P(D)}\bigg[E[C_a+C_b|D_1]P(D_1)+E[C_a+C_b|D_2]P(D_2)\bigg]\\
& =\frac{1}{P(D)}\bigg[ E[C_a|D_1]P(D_1)+E[C_b|D_1]P(D_1)+E[C_a|D_2]P(D_2)+E[C_b|D_2]P(D_2)\bigg].
\end{align*}
Let 
\[
\ep_1:= E[C~|~ 2(C+1)\geq 1/\vp]~~~\mbox{ and }~~~\ep_2:=E[C~|~ 1/\vp < 2(C+1) \leq 2/\vp];
\]
that is, $\ep_i$ is the expected number of CPs of a node that is $i$- but not $(i-1)$-vulnerable. This reduces the above to
\[
E[Z_2] = \frac{1}{\al^2+2\al\be}[(2\ep_1+\ep_2)(\al^2+\al\be) +\al\be(\ep_1+\ep_2)],
\]
which can be expressed in terms of the triangle-distribution $\{p_t\}$ as 

\begin{equation}\label{ez2}
E[Z_2] = \frac{2 \lgl tF_1 \rg}{\lgl t \rg^2}\left[ \lgl t(t-1)F_1\rg + \lgl t(F_2-F_1)\rg \left( \frac{\lgl t(t-1)F_1\rg}{\lgl t F_1\rg} +\frac{\lgl t(t-1)F_2\rg}{\lgl tF_2 \rg}\right) \right].
\end{equation}



\subsection{Transitional Distributions}\label{sec:TD}

Let $\pi_i$, $i=0,1,2$, denote the probability that a CP of an active node is of type $i$ (i.e. has $i$ active children); once we condition on the state of the parent being active (whether it's the root, in an active/inactive or an active/active pair), the states of its CPs are independent. 
In particular, the active-parent CP counts of an NP of type $i$, $(Y_{i0}, Y_{i1},Y_{i2})$, conditioned on its total number of CPs $Z_i$, is multinomially distributed with parameters $(Z_i,\pi_0,\pi_1,\pi_2)$, for $i=r,1,2$. 

To find $\pi_i$, let $C_a$ and $C_b$ be independent random variables with the size-biased triangle distribution $\{q_t\}$, and recall that a node inside the tree with $C$ CPs has degree $2C+2$ (since the parent and the linked sibling also count towards the degree). Recalling the definitions of $\al$ and $\be$ from \eqref{albedefs1} and \eqref{albedefs2}, and noting that $\pi_2$ is just $P(D)$ for the event $D$ defined in \eqref{ds}, we have
\[
\pi_1 = 2\al(1-\al-\be),~~~~~\pi_2 = \al^2 + 2 \al \be,~~~~ \pi_0=1-\pi_1-\pi_2
\]

or 
\begin{align}\label{pi1}
\pi_1 &= \frac{2}{\avt^2} \bigg( \sum^{1/2\vp} tp_t  \bigg)\bigg(  \sum_{1/\vp} tp_t   \bigg) = \frac{2\lgl tF_1\rg \lgl t(1-F_2)\rg}{\avt^2}\\
\pi_2 &= \frac{1}{\avt^2} \bigg[ \bigg(\sum^{1/2\vp} tp_t   \bigg)^2 + 2\bigg( \sum^{1/2\vp} tp_t\bigg) \bigg( \sum_{1/2\vp}^{1/\vp}tp_t \bigg) \bigg]  = \frac{\lgl tF_1 \rg^2 + 2 \lgl tF-1\rg \lgl t(F_2-F_1)\rg}{\avt^2}\label{pi2}
\end{align}

Using the distributions $p_t, r_1(t),r_2(t)$ for $Z_r,Z_1,Z_2$ (found above in \eqref{p1t},\eqref{r1t},\eqref{r2t}), we have the following:
\begin{itemize}
\item Distribution by type of CPs in generation 1 (the root is generation 0)
\[
p_1(k,l) = \sum_{t} p_t \binom{t}{k,l,t-k-l}\pi_1^k \pi_2^l (1-\pi_1-\pi_2)^{t-k-l}.
\]
\item Distribution by type of active-parent CPs of an $i$-type pair ($i=1,2$) in generation $2$ and above
\[
q_i(k,l) = \sum_t r_i(t) \binom{t}{k,l,t-k-l}\pi_1^k \pi_2^l (1-\pi_1-\pi_2)^{t-k-l}
\]
\end{itemize}
where each of these sums implicitly includes the restriction $k+l \leq t$.

Defining $R_i(z):=\sum_t r_i(t) z^t$ to be the generating functions for $r_i(t)$ defined in \eqref{r1t} and \eqref{r2t}, we have
\begin{align*}
Q_i(x,y) &= \sum_{k,l} q_i(k,l) x^k y^l \\
         &= \sum_{k,l} \sum_t r_i(t) \binom{t}{k,l,t-k-l}\pi_1^k \pi_2^l (1-\pi_1-\pi_2)^{t-k-l}x^k y^l \\
         &= \sum_t r_i(t) \sum_{k,l} \binom{t}{k,l,t-k-l}(x\pi_1)^k (y\pi_2)^l (1-\pi_1-\pi_2)^{t-k-l}\\
         &= \sum_t r_i(t) (x\pi_1+y\pi_2+1-\pi_1-\pi_2)^t\\
         &= R_i((x-1)\pi_1 + (y-1)\pi_2 +1 )
\end{align*}
and similarly
\[
P_1(x,y) = \Pi((x-1)\pi_1 + (y-1)\pi_2 +1 )
\]
where $\Pi(x) = \sum p_t x^t$ is the generating function for the triangle-distribution $\{p_t\}$. Recall that the expressions for the 
$\pi_i$ are given in \eqref{pi1},\eqref{pi2}.  As noted in \eqref{totcasc} and \eqref{fp}, the probability of a global cascade will be 
\begin{equation}\label{CascProb}
1-P_1(\bw)~~~~~~~\mbox{where $\bw$ satisfies}~~~~~\bw=\bQ\bw.
\end{equation}

\subsection{Global Cascade Threshold}

To determine whether a global cascade happens with positive probability from a single active seed, without having to explicitly calculate the probability via \eqref{CascProb}, we may refer back to \eqref{mus}, which yields
\[
\mu_{ij} = \pi_j R_i'(1) = \pi_j E[Z_i].
\]
This allows us to recast the cascade condition \eqref{cascond} in the simpler form 
\[
\pi_1E[Z_1]+\pi_2E[Z_2] >1.
\]
In Section \ref{sec:HMG} we use the formulas for $E[Z_i]$ and $\pi_i$ in \eqref{ez1}, \eqref{ez2}, \eqref{pi1}, and \eqref{pi2} to show that this condition simplifies to
\begin{equation}\label{CascadeCondition}
\frac{2}{\langle t \rangle}\bigg[ 1-\frac{\langle tF_1\rangle}{\avt}\bigg]\bigg[\langle t(t-1)F_1\rangle + \frac{\langle t(t-1)F_2\rangle \langle tF_1\rangle}{\avt -\langle tF_1\rangle}   \bigg] ~>~1.
\end{equation}
This recovers (in a totally different manner) the global cascade condition (13) from \cite{HMG}, in our special case where all edges in the network are members of triangles; networks with single edges as well as triangles were considered in \cite{HMG}. 
Even though the results in \cite{HMG} applied to the case of ``vanishingly small'' initial seed proportion $\rho_0$, they are not directly applicable to the case where there is a single (or uniformly bounded) number of seeds: an expected cascade proportion of $1$, such as in their Fig. 4, is clearly not realistic -- even in a regime that is supercritical for the spread of cascades, there will be some that die out quickly by chance  (for instance, if the seed node is not even in the giant component) and drive the expected cascade size down below 1. Rather, a sufficiently large absolute number of seeds will almost certainly ensure that a cascade will occur, in environments that are supercritical for the spread of cascades (and even in some environments that are not  -- this is the thrust of the analysis is Section II B. of \cite{HMG}). 


To explicitly check \eqref{CascadeCondition} for a specific case, consider a network whose triangle-distribution $\{p_t\}$ is Poisson with mean $\la=3$, i.e. $p_t=\e^{-3}3^t/t!$, and hence the average degree of the network is 6. To illustrate the effect of the clustering, we shall consider threshold values of $\vp=0.19$ and $\vp=0.21$, which yield the following values for the left-hand-side of \eqref{CascadeCondition}:
\beq\label{nearthreshold1}
\vp=0.19:~~~ \frac{2}{\langle t \rangle}\bigg[ 1-\frac{\langle tF_1\rangle}{\avt}\bigg]\bigg[\langle t(t-1)F_1\rangle + \frac{\langle t(t-1)F_2\rangle \langle tF_1\rangle}{\avt -\langle tF_1\rangle}   \bigg] = 6\e^{-3}+288\e^{-6}\deq 1.013>1
\eeq
\beq\label{nearthreshold2}
\vp=0.21:~~~ \frac{2}{\langle t \rangle}\bigg[ 1-\frac{\langle tF_1\rangle}{\avt}\bigg]\bigg[\langle t(t-1)F_1\rangle + \frac{\langle t(t-1)F_2\rangle \langle tF_1\rangle}{\avt -\langle tF_1\rangle}   \bigg] = 6\e^{-3}+276\e^{-6}\deq 0.983<1.
\eeq
Therefore, for this class of random networks, global cascades are possible when $\vp=0.19$, but not when $\vp=0.21$. Since all of the nodes in the networks we consider necessarily have even degree, the qualitative change across the threshold $\vp_*=\frac15$ cannot be due to the presence of nodes of degree 5 that are being activated by one active neighbor (there are no nodes of degree 5); rather, the difference must involve nodes of degree 10 being activated by two active neighbors (the parent and the sibling in the cactus). In a locally tree-like network with all edges of even degree, we would not see any change across threshold values of $1/k$ for $k$ odd -- this distinction is discussed further in the following section.

\subsection{Comparison with Unclustered Networks, Simulations}

Gleeson \cite{G1} defines the {\it extended vulnerable cluster} for locally tree-like networks to be the set of nodes that are adjacent to a large connected component of vulnerable nodes. The nodes in the extended vulnerable cluster are therefore nodes that, if chosen as the activated seed node, would trigger a global cascade;  thus, the relative size of this extended vulnerable cluster is the same as the probability that a single, uniformly randomly chosen seed will trigger a global cascade. In Appendix A of \cite{G1}, Gleeson determines analytically the relative size of the extended vulnerable cluster via the following (in \cite{W}, Watts had found this quantity numerically). Equation (A5) from \cite{G1} is 
\begin{equation}\label{GEVC}
S_e = \sum_{k=0}^\infty \tp_k [1-(1-q_\infty)^k]
\end{equation}
where $S_e$ is the relative size of the Extended Vulnerable Cluster, and $q_\infty$ is the probability that a node reached via an edge in the network is vulnerable. $q_\infty$ is found as the steady-state value of the recursion
\[
q_{n+1} = \sum_{k=1}^\infty \frac{k}{z} \tp_k [1-(1-q_n)^{k-1}]F(1,k)
\] 
where $z= \langle k \rangle$ is the mean of the network's degree distribution.

\begin{figure}
\includegraphics[scale=0.7]{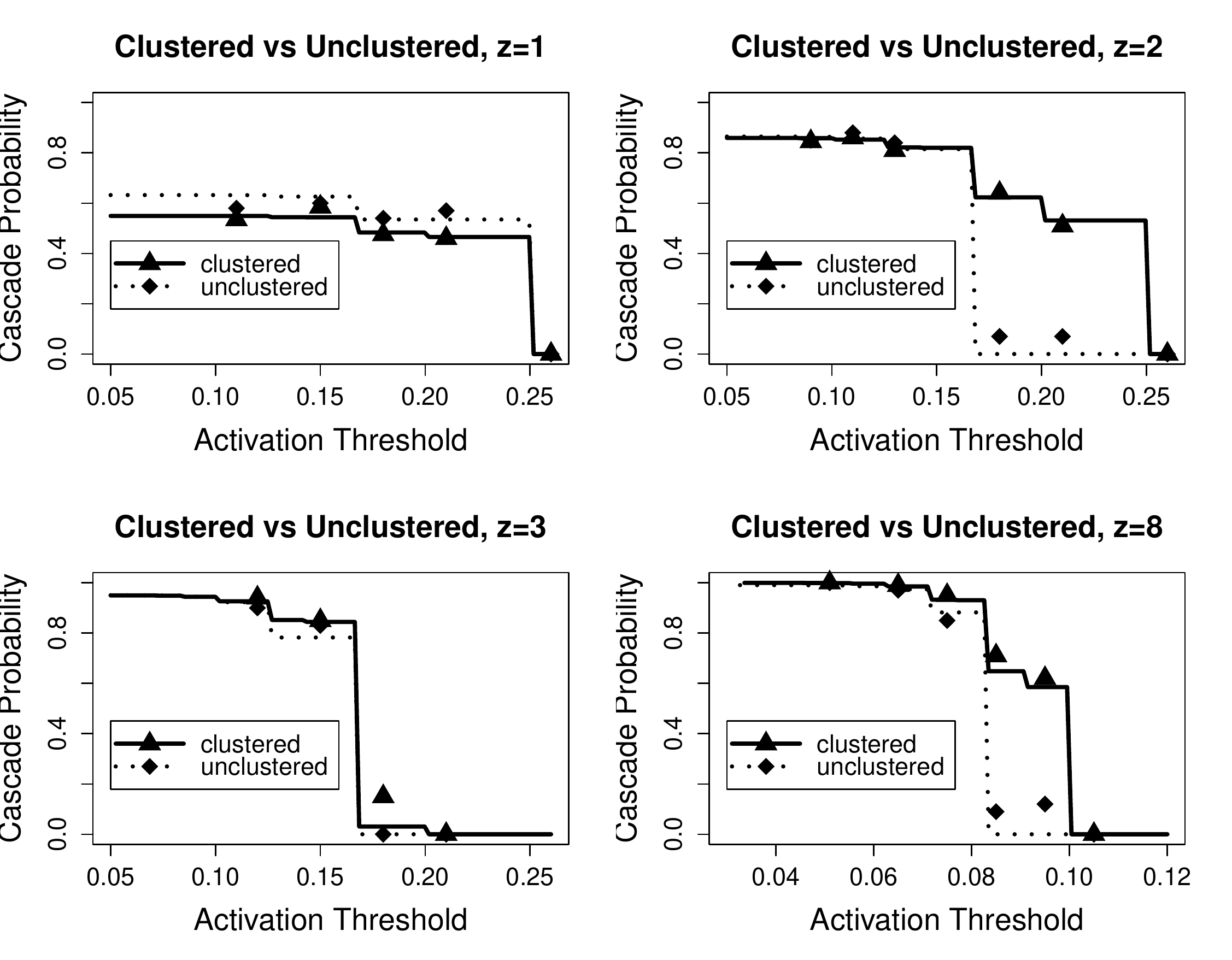}
\caption{Cascade probability versus activation threshold $\vp$ for several randomly-generated networks. Lines represent the theoretical cascade probability for clustered networks with Poisson($z$) triangle-distribution obtained from  \eqref{CascProb} (solid line), and configuration-model locally tree-like networks with the same degree distribution obtained from \eqref{GEVC} (dashed line). Points are results of simulations on networks of size 10,000 with the given degree distribution, constructed in clustered (triangles) or unclustered (diamonds) fashion.}
\label{cuplots}
\end{figure}

In Figure \ref{cuplots}, we display the cascade probability as a function of the (constant) threshold values for two networks with the same degree distributions, but where one is clustered and has Poisson($z$) triangle-distribution $p_t = \e^{-z}z^t /t!,~t=0,1,2,\dots$, and the other is a classical Molloy-Reed configuration-model network (thus is locally tree-like) with the same degree distribution $\tilde{p}_k=\e^{-z}z^{k/2}/(k/2)!~$, $k=0,2,4,\dots$. We shall refer to these networks as `clustered' and `unclustered', respectively; we perform this comparison for the values of $z=1,2,3$ and $8$.

The lines represent numerical evaluation of the cascade probabilities for the clustered network (solid line) using \eqref{CascProb}, and for the unclustered network (dashed line) using \eqref{GEVC}. Symbols represent proportions of cascades that become global, based on 200 simulations on networks of $10^4$ nodes with independently chosen seeds, for clustered (triangle) and unclustered (diamond) networks.  Note that the agreement between the simulation results and the theoretical prediction is not always sharp, but this is to be expected: as discussed in \cite{Wh}, simulation results regularly overestimate the cascade probability, since for large but finite networks, the tree- or cactus- approximation ceases to be valid in a relatively small neighborhood of the root, and the resulting loops allow cascades to take hold where they otherwise might not in a `pure' tree or cactus structure.  

Note that when the threshold $\vp$ is low, the only impediment to a global cascade is the event that  the seed is not in the graph's giant component; the case of $z=1$ illustrates that the unclustered network has a slightly larger giant component, presumably due to the absence of ``redundant'' edges that form triangles in the clustered network. However, for larger threshold values, cascades tend to be likelier in the clustered network, due to the involvement of 2-vulnerable nodes. In the case $z=3$, note the transition from nonzero to zero cascade probability around the threshold value $\vp=0.20$ for the clustered network, which is consistent with the conclusions from \eqref{nearthreshold1} and \eqref{nearthreshold2}. We also note that the jumps in the graph for the unclustered network are only at reciprocals of even integers, $1/2k, ~k=1,2,\dots$, since these are the thresholds at which a new group of nodes becomes 1-vulnerable (recall that all degrees in both networks are even). For the clustered network, since 2-vulnerability is important even in the initial spread of the cascade, we expect values of the form $2/(2k)=1/k$ to be important thresholds for $\vp$ as well; this is why we observe jumps at values of $1/k$ for $k$ odd in the clustered network but not in the unclustered one. These results are also consistent with the evidence from \cite{IHN} -- there, the authors found a window where cascades are rare in the clustered networks but nonexistent in unclustered networks.

\section{Appendix: Agreement with the Condition from \cite{HMG}}\label{sec:HMG}

In this section we flesh out the details of how our cascade condition \eqref{CascadeCondition} is equivalent to the condition in Equation (13) of \cite{HMG}. 

Recall the definitions of $\pi_j$ and $Z_i$ from Sections \ref{sec:CP} and \ref{sec:TD} respectively. $\mu_{ij} = \pi_jE[Z_i]$ means that the condition \eqref{cascond} becomes the much simpler
\beq\label{picond11}
(1-\pi_1E[Z_1])(1-\pi_2E[Z_2]) < \pi_1 E[Z_2]\pi_2E[Z_1] ~~~~\Leftrightarrow~~~~\pi_1E[Z_1]+\pi_2E[Z_2]>1.
\eeq
We have
\[
E[Z_1] = \ep_1,~~~~~~E[Z_2] = \frac{2\al}{\al^2+2\al\be}[\ep_1(\al+\be)+\ep_2 \be] 
\]
where $\ep_i$ is the expected number of CPs of an $i$-vulnerable node; in particular,
\[
\ep_1 = \frac{ \lgl t(t-1)F_1\rg}{\lgl tF_1 \rg},~~~~\mbox{and}~~~\ep_2 = \frac{ \lgl t(t-1)(F_2-F_1)\rg}{\lgl t(F_2-F_1) \rg}.
\]
Combined with the previously-found expressions for $\pi_i$, we have
\[
\pi_1E[Z_1]+\pi_2E[Z_2] = 2\al(1-\al-\be)\ep_1 + 2\al(\ep_1(\al+\be)+\ep_2\be) = 2\al(\ep_1+\ep_2\be)
\]
which, using the expressions for $\al,\be,\ep_1,\ep_2$ simplifies to 
\beq\label{e101}
\pi_1E[Z_1]+\pi_2E[Z_2] = \frac{2}{\avt}\left[ \avt \lgl t(t-1)F_1\rg + \lgl t(t-1)(F_2-F_1)\rg \right].
\eeq
In light of \eqref{e101}, our condition \eqref{picond11} is thus equivalent to Equation (13) in \cite{HMG}, suitably modified to the case of $p_{st} = \de_{s,0} p_t$. 

\section{Extensions and Conclusions}

We analyzed Watts' cascade process started with a single active seed on configuration-model-type networks with a high density of triangles. Extending Watts' work from \cite{W}, we formulated a two-type branching process to describe the initial spread of the cascade in the limit as the network size $n \to \infty$, and provided a fixed-point equation whose solution leads to the extinction probability. By numerically comparing the survival probabilities for these networks versus those for an unclustered network with the same degree distribution, we find that (i) for smaller values of the activation threshold $\vp$, the clustering decreases the global cascade probability since the clustered network is likely to have a smaller giant component, but (ii) for larger values of $\vp$, clustering increases this probability, since the presence of triangles allows the involvement of more `resistant' nodes -- ones that may require {\it two} active neighbors to activate. These nodes will in turn have higher degree than those that require only one neighbor to activate, which compounds the effects of the triangles.  

The formulas should still be valid in the case where the thresholds $\vp$ are not all equal, but instead have a nondegenerate random distribution with cumulative distribution function $F$: instead of truncating sums at $\frac{1}{2\vp}$ or $\frac1{\vp}$, we merely multiply terms by $F_i(t)$, for $i=1,2$. For the full range of networks developed by Miller \cite{Miller09} and Newman \cite{New09} (which allow single edges that are not part of triangles), it would be feasible in principle to perform a similar analysis. It would however require considering a 3-type branching process (child-pairs in states active/active, active/inactive, but also single active children); in the interest of clarity of presentation, we have not analyzed this case. It might however be instructive to analyze this case, in order to be able to interpolate smoothly between locally-tree-like networks and the highly clustered networks considered here.

\section{Acknowledgments}
We thank the participants in the SAMSI {\it Dynamics on Networks} 2010-11 working group for helpful comments and feedback.


\begin{thebibliography}{99}

\bibitem{Adler} {\sc Adler, J.} Bootstrap percolation. {\it Physica A}, {\bf 171(3),} pp. 453--470.

\bibitem{A} {\sc Amini, H.}   Bootstrap percolation and diffusion in random graphs with given vertex degrees. {\it Electronic Journal of Combinatorics}, {\bf 17,} \#R25, 2010.

\bibitem{Ar} {\sc Arthur, W.B.} Competing technologies, increasing returns, and lock-in by historical events, {\it Economic Journal}, {\bf 99,} pp. 116-131, 1989.

\bibitem{D} {\sc Durrett, R. } {\it Random Graph Dynamics}, Cambridge University Press, 2006. 

\bibitem{DCM} {\sc Dobson, I., Carreras, B.A., and  Newman, D.E.}.  A loading-dependent model of probabilistic cascading failure, {\it Probability in the Engineering and Informational Sciences}, {\bf 19 (1),} pp. 15-32, 2005.

\bibitem{F}{\sc Feller, W.} {\it An Introduction to Probability Theory and its Applications, Volume I}, 3rd ed. John Wiley \& Sons, 1950. 

\bibitem{GalCo} {\sc Galstyan, A., and Cohen, P.} Cascading dynamics in modular networks. {\it PRE} {\bf 75,} 036109, 2007.

\bibitem{G1} {\sc Gleeson, J.} Cascades on correlated and modular random networks. {\it PRE} {\bf 77,} 046117,  2008.

\bibitem{G2} {\sc Gleeson, J.} Bond percolation on a class of clustered random networks. {\it PRE} {\bf 80,} 036107, 2009.

\bibitem{GC} {\sc Gleeson, J. and Cahalane, D.}  Seed size strongly affects cascades on random networks. {\it Phys Rev. E}, {\bf 75,} 056103, 2007.


\bibitem{GMH} {\sc Gleeson, J., Melnik, S. and Hackett, A.} How clustering affects the bond percolation threshold in complex networks. {\verb arXiv ~0912.4204v2 }, 2010.

\bibitem{HMG}  {\sc Hackett, A., Melnik, S., and Gleeson, J.} Cascades on a class of clustered random networks. December 2010 (from arXiv: 1012.3651v1)

\bibitem{Hurd} {\sc Hurd, T.R., and Gleeson, J.P.} A framework for analyzing contagion in banking networks. {\tt arXiv:1110.4312}, 2011.

\bibitem{IHN} {\sc Ikeda, Y., Hasegawa, T., and Nemoto, K.}  Cascade dynamics on clustered network.  {\it 7th Intl Conf on Appl. of Phys. in Financial Analysis, J. Phys Conf series} {\bf 221,} 2010, 012005.

\bibitem{KTM} {\sc Kosterev, D.N., Taylor, C.W., and Mittelstadt, W.A.} Model validation for the August 10, 1996 WSCC system outage. IEEE Trans. Power Syst., {\bf 14(3)}, pp. 967-979, 1999.

\bibitem{mcsweeney} {\sc McSweeney, J.K.} Crossword puzzle analysis via a random network process. {\it The Mathematics of Various Entertaining Subjects: Research in Recreational Math}, Jennifer Beineke and Jason Rosenhouse, eds. Princeton University Press, 2016. 


\bibitem{Miller09} {\sc Miller, J. C.} Percolation and epidemics in random clustered networks, {\it Physical Review E}, {\bf 80,}, 020901, 2009.

\bibitem{MR} {\sc Molloy, M., and Reed, B.} A critical point for random graphs with a given degree sequence. {\it Rand. Struct. Alg.} {\bf 6,} 161,  1995.

\bibitem{New09} {\sc Newman. M.E.J.}  Random graphs with clustering, {\it Phys. Rev. Lett.} {\bf 103,} 058701, 2009. 

\bibitem{W} {\sc Watts, D.} A simple model of global cascades on random networks. {\it PNAS} {\bf vol 99 no 9} pp. 5766-5771, 2002.

\bibitem{Wh} {\sc Whitney, D.} A dynamic model of cascades on random networks with a threshold rule.  Available at {\tt arXiv:0911.4499}, 2009.

\end{thebibliography}
\end{document}